# High-precision FoM measurement setup for Ti:Sapphire crystals

**Vojtěch Miller**[1,3*] **and Karel Žídek**[2]

[1] *CRYTUR, spol. s r.o., Na Lukách 2283, 511 01 Turnov, Czech Republic*
[2] *TOPTEC, Institute of Plasma Physics of the Czech Academy of Sciences, Za Slovankou 1782/3, 182 00 Prague 8, Czech Republic*
[3] *Technical University of Liberec, Studentská 1402/2 461 17 Liberec 1, Liberec, Czech Republic*

*\*vojtech.miller@tul.cz*

**Abstract:** The figure of merit (FoM) of Ti:Sapphire (Ti:Sa) crystals is a generally used means to evaluate the quality of the crystals. Despite the importance of Ti:Sa, the question of FoM measurement precision stayed out of focus, while the commercially available spectrometers provide unsatisfactory 3σ precision reaching ±60 %. In this paper, we present a setup for a single-pass high-precision transmission measurement for three different wavelengths (532 nm, 780 nm, and 1560 nm) based on Nd:YAG and Er:YAG lasers. A synchronous detection via a double integrated sphere enabled us to achieve the transmission uncertainty of 0,01-0,03%. With the presented setup, we show that it is possible to determine the FoM values with 3σ precision of ±7,5 %. Owing to the high FoM precision, we were able to trace spatial inhomogeneities of an unannealed Ti:Sa crystal produced by a commercial manufacturer Crytur. Our measurements demonstrate that the FoM values can be significantly affected by the crystal inhomogeneities and angular mismatch between the c-axis of the Ti:Sa and polarization orientation.

## 1. Introduction

Ti:Sapphire (Ti:Sa) is an exceptional material widely used as an infrared lasing medium [1–3]. The advantages of Ti:Sa are in the tunability of the lasing elements in the extensive range of 600-1100 nm. The large lasing range is a fundamental requirement for using Ti:Sa lasers to generate ultra-short pulses (femtosecond scale) [4, 5].

Ti:Sa material quality is evaluated through the parameter denoted as a figure of merit (FoM). The FoM is defined as the ratio of absorption coefficients in the pumping region divided by a residual absorption in the lasing region. The exact wavelengths are not determined and depend on the particular lasing and pumping wavelengths attributed to the application of interest. Commonly accepted wavelengths are 514 or 532 nm for the pumping region and 800±50 nm for the lasing region [6–9]. The reason behind the importance of FoM lies in the parasitic reabsorption in the lasing region. This unwanted reabsorption of light leads to effective losses during lasing and increases the total thermal load of the gain medium. The high thermal load can lead to local overheating and destruction of the active medium.

FoM measurements have been reported in several articles. However, none of them discusses the real measurement precision of the FoM parameter and it is, therefore, difficult to judge the actual quality of the crystals [7, 10–13]. This is one of the reasons why the actual quality of the laser crystal and its FoM are commonly acquired through the direct monitoring of the crystal in a laser cavity during the laser operation. This is, naturally, the most trustworthy evaluation of the crystal. At the same time, this procedure requires the fabrication of a high-quality laser element and its alignment in the cavity, while it would be immensely useful to evaluate the quality on a small plan parallel crystal element, which is incomparably simpler to manufacture.

Another issue lies in the number of parameters affecting the laser element qualification. The critical factors include the variation of FoM across the crystal, deviation of rotation of the c-axis to the electric field of the incident light wave, cleanliness of the surface, and purity of the bulk material. Any volume defects (bubbles, fog, impurities) cause scattering inside the crystal, affecting the transmittance measurement. These factors have been consistently neglected in the published literature [13–15]. This makes the evaluation of FoM even less reliable.

In this paper, we address the issue of high-precision FoM measurement via single-pass measurement. We study FoM both in the light of reaching high-precision transmission measurement, as well as gaining knowledge about critical factors – namely crystal inhomogeneities and orientation.

We demonstrate that it is possible to reach a remarkably high precision of the measurements using Nd:YAG and Er:YAG lasers combined with synchronous detection via an optical dual frequency chopper. The presented setup and the low error of measurement allowed us to reliably trace inhomogeneities in the crystal, as well as the effect of crystal rotation. We demonstrate these measurements on an unannealed Ti:Sa crystal with prominent local variation in the optical properties. We discuss the role of each factor in the measurement.

With the presented setup, we show that it is possible to determine FoM via single-pass measurement with a precision (standard deviation) of 2,5 %. In comparison, the uncertainties introduced in the commonly used measurements limit the precision to 20 %. To provide a deeper insight -- commonly used commercial spectrometers typically reach a precision of 0,1 %. Should we measure a 10 mm thick Ti:Sa sample with absorption of 2 $cm^{-1}$, we will reach the absorbance of 0.01 $cm^{-1}$ at 780 nm wavelength for FoM 300. As a result, the total absorption within the sample at this wavelength would be less than 0,2 %. Therefore the relative error of such measurement would be unacceptably high. On the contrary, samples with a high absorbance at excitation wavelength rapidly approach here 100%, and it is very difficult to measure the absorption coefficients here reliably.

Therefore, this article demonstrates a possible way to construct a single-pass high-precision transmittance measurement. At the same time, the results can serve as a guideline to gain a low-error FoM measurement.

## 2. Experimental setup

Due to insufficient precision achieved for FoM evaluation by commercially available instruments, a new laser-based experimental setup has been built. The necessity for higher precision lies in the requirements for determining high FoM material, where the absorption coefficients in the excitation and lasing spectral regions are different by two orders of magnitude. It is very demanding to measure them both with reasonable precision.

The presented setup was based on three distinct wavelengths produced from two lasers - second harmonic generation (SHG) of Nd:YAG laser (532 nm) together with fundamental beam and SHG of Er:YAG laser (780 and 1560 nm). In particular, we created a single-pass multi-wavelength high-precision measurement of transmittance. The high precision was attained by combining a high-intensity light source, lock-in detection, and high dynamic range measurement of intensity, where the detection efficiency was independent of the position of the incident beam.

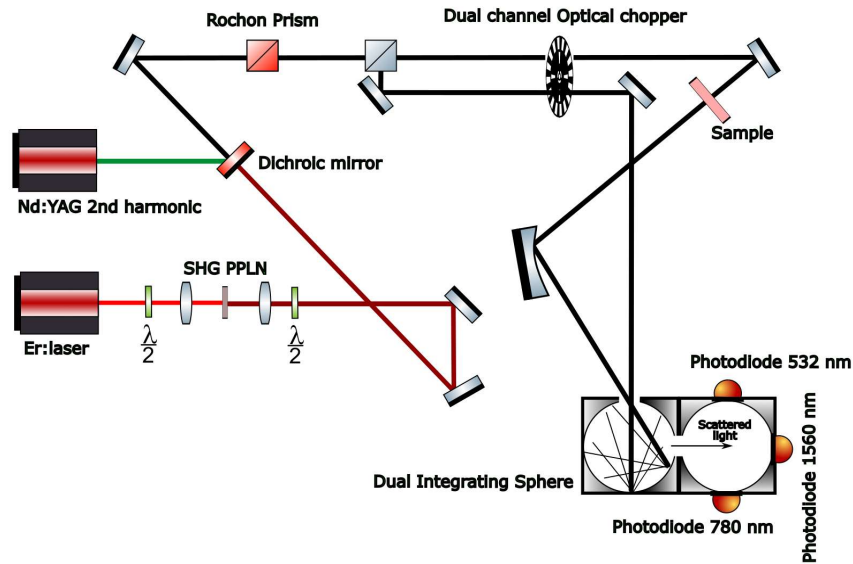

Figure 1: Dual laser-based experimental setup scheme for high-precision transmission measurement on 532 nm, 780nm, and 1560 nm.

The laser beams were taken from two input branches aligned using a dichroic beam splitter Thorlabs DMLP650. The first branch consisted of Nd:YAG laser, where we utilized the SHG, while the IR filter was used to suppress any residual 1064 nm light. The 532 nm laser was preferentially polarized horizontally with a low degree of polarization. The second input branch consisted of Er:YAG laser Toptica Photonics FemtoFErb 1560, which was used on its fundamental wavelength of 1560 nm and combined with its second harmonic frequency (780 nm) by an SHG crystal (PPLN, Covesion). Prior to the SHG, the Er:YAG beam polarization was adjusted via λ/2 waveplate for 1560 nm (Thorlabs WPH05M-1560) and focused through a lens to the SHG crystal. Subsequently, it was recollimated. SHG and fundamental beams were transmitted through a λ/2 waveplate for 780 nm (Thorlabs WPH05M-780) so that the 780 nm beam was horizontally polarized, while the fundamental beam features approximately circular polarization.

All three beams propagated along the same beamline through the rest of the setup. First, we aimed to set their polarization with high precision to purely horizontal polarization. Therefore, we used a Rochon prism Thorlabs RPV10 to split the horizontal and vertical degree of polarizations with a ratio of $1:10^5$. Such a degree of polarization is a prerequisite for the measurement since the dependence of the transmission of the Ti:Sa on the polarization would otherwise distort the results. The angle of the Rochon prism was adjusted based on the minimization of reflection from a sample at Brewster angle.

The horizontally polarized beams were then divided into two separate branches by a beam splitter. One branch (denoted as a sample branch) was transmitted through the sample. The other branch (denoted as a reference branch) was used to compensate for the fluctuations of the intensity of the lasers. The sample and reference beams passed through the dual channel (ratio 3:7) optical chopper blade to modulate their intensity.

Both sample and reference beams were detected by a pair of integrating spheres, where the intensities were detected by three separate photodiodes dedicated for each wavelength -- 2x Thorlabs DET36A2 for 532 and 780 nm and 1x NewPort Large Area Photodetector Model 2317 for 1560 nm. The spectral selectivity was attained by using the sensitivity of the photodiode combined with a series of color filters. The pair of integrating spheres was used to avoid any distortion connected with a shift in the detected beam due to the fact that the efficiency of a photodiode is spatially inhomogeneous, i.e. each spot features a different sensitivity because a single integrating sphere has proven not to reach a satisfactory level of homogenization. A neutral density filter was used in the sample branch for the measurements at 1560 nm to attain a comparable intensity of the reference and sample beam.

The light in the sample branch propagated through the sample, where it was partially absorbed, and finally, it was reflected by an off-axis mirror to the pair of integrating spheres. The signal from the photodiodes was then amplified and digitally processed via Fourier transformation to extract the sample and reference beam intensity. An essential advantage of the setup was that we could use the same detector to detect both sample and reference beam intensity since they could be traced based on their modulation via an optical chopper.

The samples were mounted on the tip, tilt, rotation stage Thorlabs TTR001/M that allowed complete alignment of the sample. The alignment of the sample was one of the investigated points of this paper. The stage was further mounted on a Thorlabs motorized 1D stage used to run a lateral scan through the samples. The ratio of sample and reference sample for the motor position with the sample beam not hitting the sample was used as 100% reference. This reference was measured before and after each spatial scan to avoid long-term drift in the values.

**Achievable precision**

Initially, we measured the long-term stability of the detected transmission without any inserted sample. The long-term stability has been calculated from several thousand values corresponding to several hours of continuous measurement. The calculated precision in Table 1 corresponds to the standard deviation (STD) over a studied temporal window. We applied the period of 20 seconds, which corresponds to the duration of a single point measurement, as well as 180 seconds, which corresponds to the duration of a 1D scan through the sample.

Table 1: Achieved precision expressed as a standard deviation (STD) on wavelengths for the time scales of 20 and 180 s for individual wavelengths.

| Wavelength [nm] | Mean STD (20s) | Worst case STD (20s) | Mean STD (180s) | Worst case STD (180s) |
| --- | --- | --- | --- | --- |
| 532 | 0,015 % | 0,050 % | 0,029 % | 0,050 % |
| 780 | 0,026 % | 0,060 % | 0,029 % | 0,045 % |
| 1560 | 0,007% | 0,018 % | 0,020 % | 0,050 % |

When performing a 1D lateral scan of a sample, the precision was further enhanced by double measuring the reference value at the beginning and at the end of the measurement. The resulting transmission was compensated accordingly.

### 3. Reference measurement

To carry out the test of the device, we prepared a set of four samples from a low-dispersion glass with thicknesses of 1, 2, 4, and 8 mm. The samples were manufactured using the same realiable and well reproducable technological procedure from a single bulk piece of glass to avoid any differences between the samples apart from the thickness. Therefore, the samples featured the same Fresnel reflectance, absorption coefficients, and scattering connected to the surface imperfections. Based on this, we could verify that the transmission through the samples on each wavelength follows the Lambert-Beer law.

Due to the high precision of the transmission measurement, the quality of the polishing procedure has become one of the crucial factors. Different glass or sapphire substrates, including the commercially available ones, can feature a decreased transmittance due to their surface quality. While this effect is not important for the standard precision of commercially available spectrometers, in our case it posed an issue,

The samples were measured on each of the wavelengths with a 0° incident angle. We measured a series of 19 spots in a lateral 1D scan across each sample where each spot was 1 mm apart from each other – see Figure 2. The variation of local transmittance across the sample was more pronounced than the measurement error itself. This variation can be attributed to the local impurities in the material, remaining dust particles, the role of polishing and other local defects. We created a procedure to extract a single value from a spatial scan along the sample. In particular, we used the mean value within the series. The spatial scan was repeated ten times to attain a realistic estimate of statistical precision for each spot.

The overall transmittance measured for four different glass thicknesses – see Figure 3 – allowed us to fit the data with an exponential function based on the Lambert-Beer law. The amplitude of the fit corresponded solely to Fresnel losses connected to the refractive index of the material. The error was given by the measurement error itself (scale of 0.02 %) and the sample inhomogeneity (scale of 0.15 %). The process described above has been done for all three wavelengths. The results were then plotted in Figure 3 below.

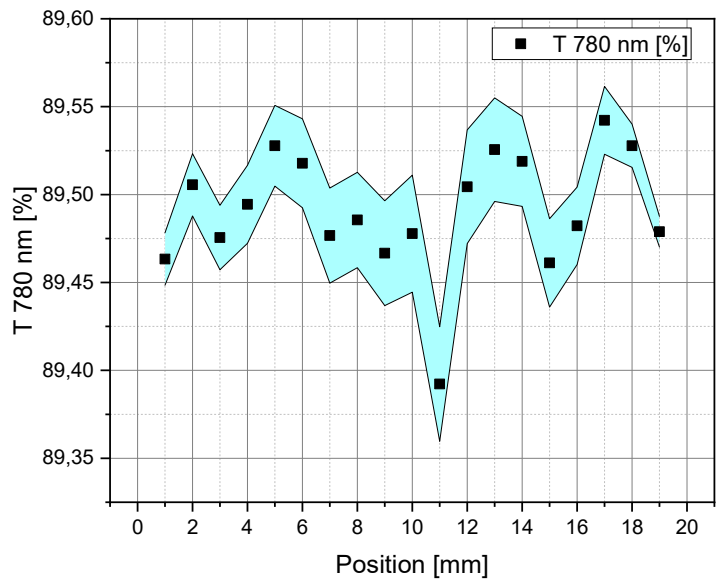

Figure 2: Local transmittance along 1 mm glass sample measured for 780 nm wavelength. Lateral scan with 1 mm step. Each point represents the mean from six subsequent measurements on the spot. Shaded area represents STD region of the measurement.

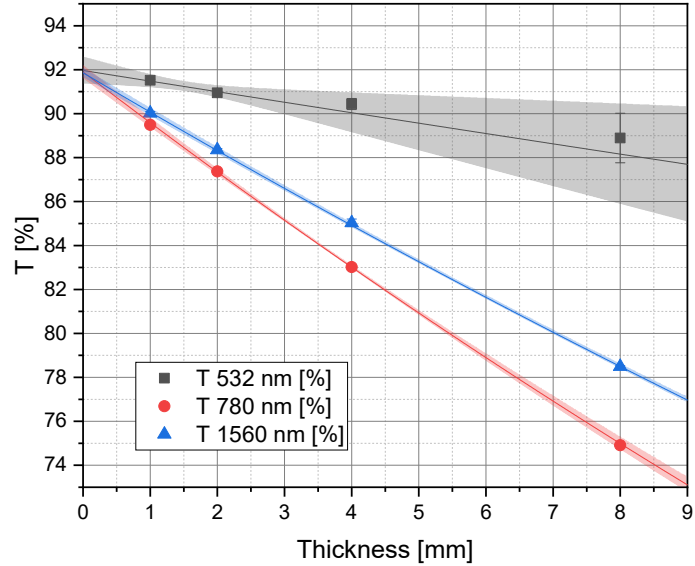

Figure 3: Test measurements of transmittance for glass test samples of gradually increased thickness 1-8 mm. Each value was measured by respective wavelength and is calculated as a mean from twenty scanned points in one axis with a distance of 1 mm between each scan.

The results of the fitted amplitude for three wavelengths are listed in Table 2, together with the estimate of fitting parameters errors provided by Origin software. –The results followed, within the experimental error, the reference results measured on the commercial spectrometer Photon RT – see Table 2

Table 2: Extrapolated values of transmission from data shown in Figure 3, the corresponding STD, and index of refraction calculated from the Fresnel equation. The last two columns represent comparable extrapolated data based on a commercial spectrometer.

| Wavelength [nm] | T Exp [%] | STD Exp [%] | Index of refraction | STD Index of refraction | T Spectr [%] | STD Spectr [%] |
|---|---|---|---|---|---|---|
| 532 | 91,83 | 0,14 | 1,514 | 0,006 | 91,6 | 0,3 |
| 780 | 91,90 | 0,07 | 1,511 | 0,003 | 91,7 | 0,2 |
| 1560 | 91,86 | 0,05 | 1,512 | 0,002 | 91,6 | 0,4 |

To provide another consistency check of the results, we estimated with high precision the refractive index of the glass based on the measured transmittance and Fresnel equations. Consequently, by using an independent method, we were able to derive a consistent value. Namely, we used an interferometer Trioptics OptiSurf LTM to measure the optical path on the 8 mm thick glass sample. By using an independent tool to measure the thickness of the glass, we were able to calculate the material refractive index. Since the interferometer used the wavelength of 1310 nm for the optical path, the resulting value of the refractive index was valid for this wavelength, and it reached 1,508. This is in very good agreement with the estimated refractive index from the Fresnel equation, where the values for 780 nm and 1550 nm were expected to be nearly equal to the one at 1310 nm.

## 4. Ti:Sa FoM measurement

The core of our work lied in the high-precision determination of FoM value for Ti:Sa samples. Therefore, we turned to a thorough study of this type of measurement.

**FoM calculation**

The FoM of Ti:Sa was calculated as a proportion of the absorption coefficients on 532 and 780 nm wavelengths:

$$FoM = \frac{\alpha_{532nm}}{\alpha_{780nm}} \tag{1}$$

The absorption coefficients were calculated from the measured transmission and known Fresnel reflection coefficients for the zero angle of incidence:

$$\alpha_i = \frac{\ln\left(\frac{\left(1 - R_{p_i}\right)^2 + R_p^2}{T_i}\right)}{d}, \tag{2}$$

where $\alpha_i$ was absorption coefficient on the respective wavelength, $R_{p_i}$ is the Fresnel reflection coefficient of sapphire on the particular wavelength from Ref. [16], and d is sample thickness. The model considers reflections from the 1st interface, the 2nd interface, and the second-order internal reflection from the 2nd interface. For sapphire, the higher order reflections reach the level of $10^{-4}$ % and can be, therefore, neglected. The zero angle of incidence was set with high precision (< 0.2 degrees) by observing the back-reflection of the laser beams.

**Preparation of Ti:Sa samples**

Measurement on real-life samples was performed using Ti:Sa material provided by Crytur comp., a commercial producer of monocrystalline materials. The presented results were attained on a polished sample 2.54 mm thick with a diameter of 37 mm from unannealed Ti:Sa – see Figure 4 for the photo of the sample. The material was grown by the Czochralski method. The sample had a concentration of $Ti_2O_3$ reaching 0,1 %. The orientation of the sample was in the a-axis, i.e., the polished sample front is normal to the a-axis.

**Spatial scan of Ti:Sa samples**

Since the unannealed sample of Ti:Sa suffers from prominent spatial inhomogeneities in the absorbance, we used the sample to study the spatial dependence of FoM, which is in the literature commonly provided as a single value for a crystal, despite the inhomogeneities.

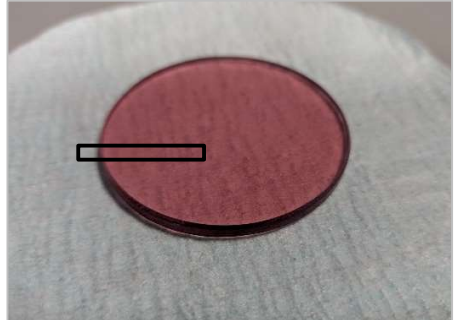

*Figure 4: The studied Ti:Sa sample with a marked area where the lateral 1D scan took place. The data at the very edge of the sample were omitted until the stable output was reached.*

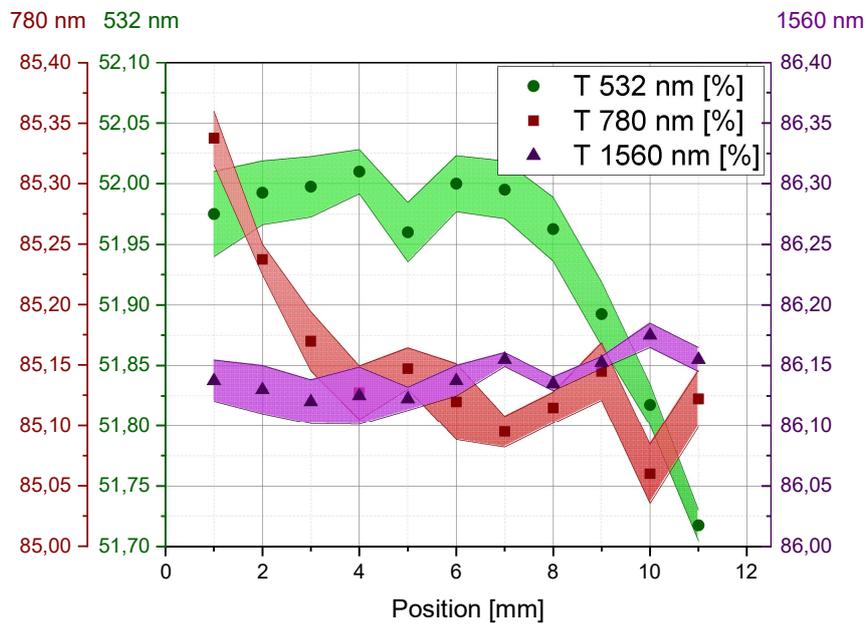

Figure 5: Local transmission on Ti:Sa sample on the three studied wavelengths measured for the perpendicular incidence, electric field oriented parallel to the axis c. A lateral scan was carried out with a 1 mm step. Each point represents the mean from five independent measurements on the same spot, where the error is determined as the STD from the value. Note that each wavelength uses a different y-axis scaling to provide a better comparison. The shaded area represents the STD region of the respective measurement.

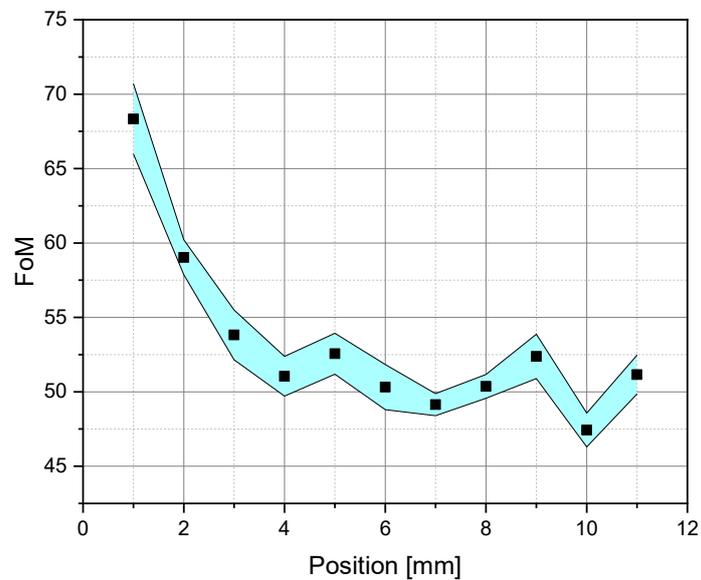

Figure 6: Local distribution of the FoM parameter derived from data in Figure 4 via equations (1) and (2). Each point represents the mean value from five independent measurements on the same spot. The error value is calculated as the STD of the five FoM values. The shaded area represents the STD region of the measurement.

On Figure 5 are depicted the local variation in transmission across the Ti:Sa sample when scanned along the c-axis. The figure clearly shows that the precision of the measurement is high enough to trace the local profile of the transmission inside the crystal. The achieved precision is in order of 0.05 %, whereas the local impurities feature changes up to 0.25%. The FoM values were calculated based on the transmission (absorption coefficient respectively) on wavelengths 532 nm and 780 nm, as described previously. The measurement on the 1560 nm is present as a reference measurement to check whether there are any local impurities that are of different nature than the atomic state of Ti ions, such as local inhomogeneity, microbubbles, or fog.

The FoM values have not reached high numbers for the presented sample, which was expected since we studied a non-annealed sample. The key benefit of the graph is to see the achievable precision of the FoM measurement, which is 1.4. The relative deviation of FoM is then 2.5 % (7.5 % for 3 sigma). This is a major improvement over the standard spectrometer measurements available on the market. As an example can be used spectrometer Photon RT which has a measurement accuracy of transmission of 0,1%. Such deviation will result in FoM 3 sigma deviation of almost 60 %, i.e., an order of magnitude higher.

At the same time, it can be spotted in Figure 6 that FoM varied across the studied sample between 47 and 67, i.e., the number can be different by as much as 30 % between different spots.

**Rotation of the Ti:Sa crystal**

Another source of inaccuracy present in the measurements of FoM is the deviation of the c-axis of the crystal lattice of Ti: Sa from the polarization of the incident light. There is a significant dependence of the transmission through the Ti:Sa material along its axis [11, 14, 17]. The light polarized along the c-axis has roughly double the absorption coefficient against the m-axis or the a-axis. Therefore, the inaccuracy of the orientation of the sample will result in different transmission and FoM values.

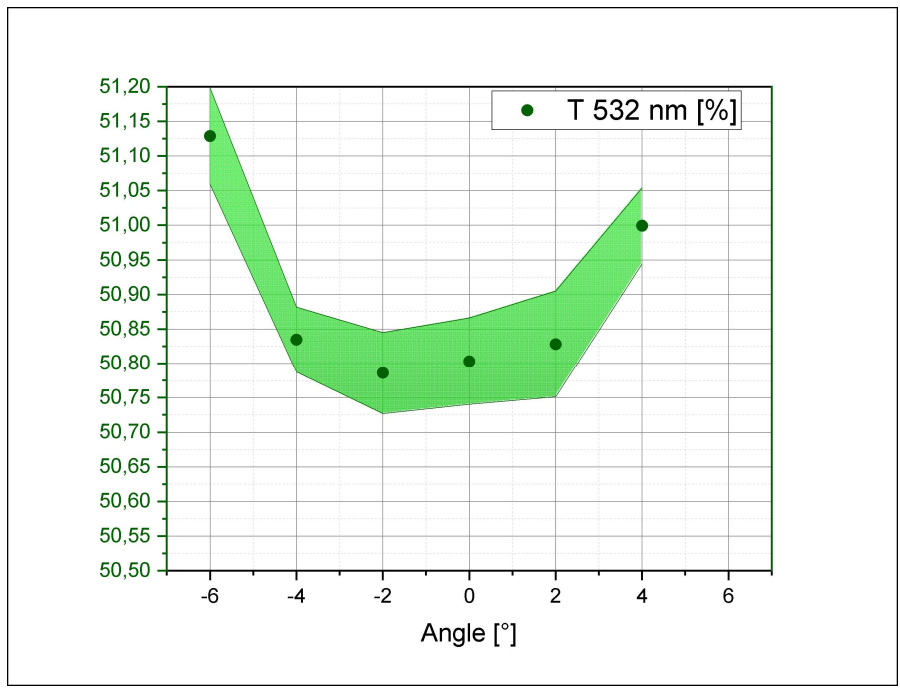

Figure 7: Transmission dependence on the angular orientation of the Ti:Sa c-axis to the electric field orientation. Each point represents the mean value from seven independent measurements on the spot. The shaded area represents the STD region of the measurement.

The dependence of transmission on the rotation of the crystal axis is illustrated in Figure 7. The figure shows the results for the wavelength of 532 nm, as here, the angle dependence is highly prominent. The tip, tilt, rotation stage has been gradually tilted by 2 degrees. Tilting the sample led to a change in the measured position on the sample, which was partly compensated by shifting the measured data; still, it is worth noting that the measured points were not identical. Nevertheless, the strong tilt angle-dependent absorption at 532 nm can be traced, following cosine dependency with the rotation of the c-axis.

Since the measured points in Figure 7 do not perfectly align when rotated, the STD of the measured data is higher compared to the data in Figure 4. Here, in addition, the error is increased by the local variation of the transmission across the sample.

In previous sections (Figure 5), we showed that the presented setup can determine FoM values with the STD of 2,5 %. To provide a comparison to the error caused by sample rotation, we calculated that the rotation of a sample by 3 degrees will introduce the relative error of absorbance reaching 0,2 %, which will transpose into the change in FoM by 3 %. This leads to the conclusion that the sample must be oriented correctly in the measurement setup well below ±3° not to affect the overall measured values. These values demonstrate the need for fine-tuning crystal angle for any high-precision FoM measurement.

## 5. Conclusions

We carried out a thorough study addressing the evaluation of Ti:Sa quality. We have been able to successfully build an experimental setup allowing us to measure transmittance with precision up to 0,01 %, which allows us to calculate the FoM of the Ti:Sa material with a relative standard deviation of 2,5 %. The correctness of the measurement has been concluded on a set of glass samples with the index of refraction n=1,513. We have also been able to show the influence of rotational misplacement of the sample and deviation of the c-axis to the E-field of the beam, respectively.

The presented setup allows for efficient measurement of the FoM prior to manufacturing laser-grade element, which brings invaluable benefits to the attempts to study the process of Ti:Sa crystal growth and the governing factors of the process. It also provides insight into the quality of the bulk material from the spectroscopic view.

**Funding.** Ministry of Education, Youth and Sports ("Partnership for Excellence in Superprecise Optics," Reg. No. CZ.02.1.01/0.0/0.0/16_026/0008390).

This work was partly supported by the Student Grant Scheme at the Technical University of Liberec through project No. SGS-2022-3083.

**Acknowledgments.** We thank Jana Preclíková for the initial idea to start a project dealing with the precision of the FoM evaluation on Ti:Sa material.

**Disclosures.** The authors declare no conflicts of interest.

**Data availability.** Data underlying the results presented in this paper are not publicly available at this time but may be obtained from the authors upon reasonable request.